\documentclass[aps,prl,reprint,superscriptaddress]{revtex4-1}
\usepackage[usenames,dvipsnames]{color}
\usepackage{amsmath,array,amssymb}
\usepackage{graphicx}
\usepackage{hyperref}
\usepackage{textcomp}
\usepackage{braket}
\usepackage{multirow}
\usepackage{indentfirst}
\usepackage[FIGTOPCAP,raggedright,nooneline,bf,footnotesize]{subfigure}
\usepackage{paralist}
\usepackage{overpic}

\renewcommand{\vec}[1]{\boldsymbol{#1}}

\renewcommand{\eqref}[1]{Eq.~(\ref{#1})}

\begin{document}
\title{Quadratic Zeeman effect and spin-lattice relaxation of Tm$^{3+}$:YAG \\ at high magnetic fields}

\author{Lucile Veissier}
\email{lveissie@ucalgary.ca}
\affiliation{Institute for Quantum Science and Technology, and Department of Physics \& Astronomy, University of Calgary, 2500 University Drive NW, Calgary, Alberta T2N 1N4, Canada}

\author{Charles W. Thiel}
\affiliation{Department of Physics, Montana State University, Bozeman, Montana 59717, USA}

\author{Thomas Lutz}
\affiliation{Institute for Quantum Science and Technology, and Department of Physics \& Astronomy, University of Calgary, 2500 University Drive NW, Calgary, Alberta T2N 1N4, Canada}

\author{Paul E. Barclay}
\affiliation{Institute for Quantum Science and Technology, and Department of Physics \& Astronomy, University of Calgary, 2500 University Drive NW, Calgary, Alberta T2N 1N4, Canada}

\author{Wolfgang Tittel}
\affiliation{Institute for Quantum Science and Technology, and Department of Physics \& Astronomy, University of Calgary, 2500 University Drive NW, Calgary, Alberta T2N 1N4, Canada}

\author{Rufus L. Cone}
\affiliation{Department of Physics, Montana State University, Bozeman, Montana 59717, USA}

\begin{abstract}
Anisotropy of the quadratic Zeeman effect for the $^3{\rm H}_6 \rightarrow \, ^3{\rm H}_4$ transition at 793 nm wavelength in $^{169}$Tm$^{3+}$-doped Y$_3$Al$_5$O$_{12}$ is studied, revealing shifts ranging from near zero up to + 4.69 GHz/T$^2$ for ions in magnetically inequivalent sites. This large range of shifts is used to spectrally resolve different subsets of ions and study nuclear spin relaxation as a function of temperature, magnetic field strength, and orientation in a site-selective manner. A rapid decrease in spin lifetime is found at large magnetic fields, revealing the weak contribution of direct-phonon absorption and emission to the nuclear spin-lattice relaxation rate. We furthermore confirm theoretical predictions for the phonon coupling strength, finding much smaller values than those estimated in the limited number of past studies of thulium in similar crystals. Finally, we observe a significant -- and unexpected -- magnetic field dependence of the two-phonon Orbach spin relaxation process at higher field strengths, which we explain through changes in the electronic energy-level splitting arising from the quadratic Zeeman effect.

\end{abstract}

\maketitle
\newpage

\section{Introduction}
Tm$^{3+}$:Y$_3$Al$_5$O$_{12}$ (YAG) has been intensely studied and widely used for applications ranging from photonic signal  processing \cite{cole_coherent_2002,lavielle_wideband_2003} and laser frequency stabilization \cite{strickland_laser_2000} to solid-state quantum memories \cite{lauro_slow_2009,bonarota_efficiency_2010}. Indeed, the optical $^3{\rm H}_6 \rightarrow \, ^3{\rm H}_4$ transition of this material exhibits desired properties, such as long coherence lifetimes of up to 105 $\mu$s in zero field \cite{macfarlane_photon-echo_1993} and 300 $\mu$s in a small magnetic field \cite{thiel_measuring_2014}. Furthermore, its wavelength of 793 nm is easily accessible with commercial laser technologies. Another advantage is the absence of hyperfine structure (and corresponding spin-flip sidebands) in zero magnetic field for the single thulium isotope $^{169}$Tm, which has a nuclear spin of $1/2$ and therefore no nuclear quadrupole structure; the absence of sidebands enhances resolution in optical-microwave frequency spectrum analysis applications. An external magnetic field lifts the degeneracy of spin states through the enhanced effective nuclear Zeeman interaction \cite{abragam_enhanced_1983}, and optical pumping of those nuclear hyperfine states provides long population storage lifetimes and enables many spectral hole burning applications \cite{macfarlane_spectral_1987,ohlsson_long-time-storage_2003,goldner_hyperfine_2006,louchet_branching_2007}.

\begin{figure}[t]
\centering
\includegraphics[width=0.8\columnwidth]{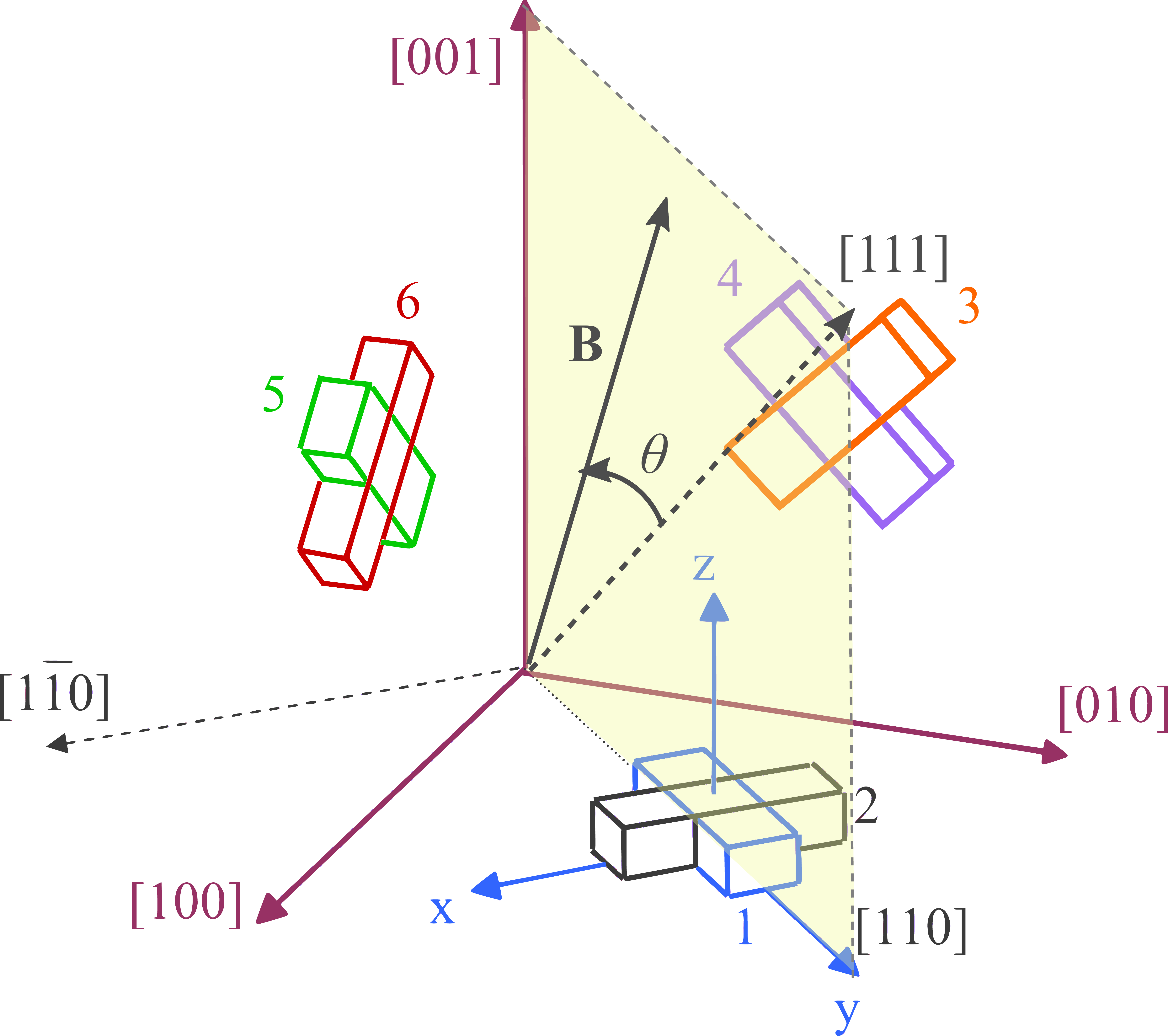}
\caption{Orientation of the six different sites of Tm$^{3+}$ ions in the YAG crystal lattice \cite{dillon_ferrimagnetic_1961,tongning_ralentir_2014}. Each parallelepiped represents the orientation of the local $D_2$ point symmetry for a subset of sites. The specific set of $x$, $y$, and $z$ axes labeled in the diagram correspond to the local frame of site 1. In our experiment, the light propagates along $[1\bar{1}0]$ and the magnetic field is applied in the plane defined by $[110]$ and $[001]$, forming an angle $\theta$ with the $[111]$ axis. Because of the $D_2$ point symmetry of the sites, the electric transition dipole moment for the specific ground and excited states involved in the 793 nm optical transition is aligned along the local $y$ axis. The gyromagnetic tensor $\gamma_J$ orientation is also determined by the local frame.}
\label{fig:yag}
\end{figure}

In order to find optimal operation parameters for applications, characterization of the orientation and site dependent properties such as spin-state lifetimes and magnitude of the quadratic Zeeman shift is required. This is particularly important in the case of Tm:YAG, since due to the cubic symmetry of the YAG crystal structure, the lattice contains six classes of thulium sites that have the same point symmetry but different local orientations relative to externally applied electric and magnetic fields \cite{sun_symmetry_2000} (see Fig. \ref{fig:yag}). Because of a large anisotropy in the magnetic properties of the thulium ions, many spectroscopic properties depend strongly on the orientation of the applied magnetic field and, as a result, different behavior can be observed for ions at each of the six magnetically-inequivalent sites. In particular, this site dependence has been observed for the thulium nuclear spin-state lifetimes \cite{louchet_branching_2007}. In addition, the quadratic Zeeman effect, caused by the mixing of the crystal field levels due to the applied magnetic field, shifts the energy levels in both the ground and excited states, leading to a shift of the optical transition that is expected to be strongly orientation and site dependent. Finally, by observing the properties at higher magnetic field strengths, physical mechanisms with different field sensitivities may be separated and unambiguously identified, improving the general understanding of material properties.

This paper is arranged as follows. After outlining the theoretical background, the orientation dependence of the optical transition frequency is experimentally characterized and compared to the theoretical prediction. We also measure the spin-lattice relaxation of the thulium sites with the longest hyperfine lifetime as a function of the magnetic field strength and temperature. We identify the contribution from the direct phonon process, finding that it only becomes larger than the two-phonon Orbach relaxation process for fields greater than 4 T - this holds even at temperatures below 2 K. These results reveal that the phonon coupling strength in Tm:YAG is much smaller than what has been previously estimated for thulium in other hard oxide crystals, suggesting that past measurements \cite{suzuki_squid_1980,suzuki_enhanced_1981,roinel_relaxation_1985} were limited by other relaxation mechanisms such as paramagnetic impurities. 

\section{Quadratic Zeeman shift}
\subsection{Theoretical background}
When studying properties of rare-earth transitions at higher external magnetic field strengths (generally $B> 1$~T), one must consider the second-order mixing of crystal-field wavefunctions due to the quadratic Zeeman effect. Here we study this effect in order to be able to predict the position of energy levels and transition frequencies at different magnetic fields up to 6 T. Furthermore, as we will show later, in the case of Tm:YAG the anisotropy of the quadratic Zeeman effect allows us to spectrally resolve different subsets of ions.

For a Tm$^{3+}$ ion in a crystal matrix, the Hamiltonian is given by
\begin{equation}
H = H_{\rm FI} + H_{\rm CF} + H_{\rm HF} + H_{\rm Q} + H_{\rm eZ} + H_{\rm nZ} \, ,
\label{eq:Htotal}
\end{equation}
where the first two terms are the largest, corresponding to the free ion and the crystal field coupling. Consequently, the wavefunctions of the system can be expressed by the eigenstates of these two terms together, and the other terms, namely the hyperfine, nuclear quadrupole, electronic and nuclear Zeeman interaction, are treated as perturbations \cite{macfarlane_coherent_1987}. The Tm$^{3+}$ ion is a non-Kramers ion that sits at a site with $D_2$ point symmetry in the YAG lattice, resulting in a singlet electronic ground state. Consequently, the electronic angular momentum $J$ is effectively quenched to first order and the hyperfine structure is given by second-order terms \cite{teplov_magnetic_1968}. Taking into account the $^{169}$Tm nuclear spin of $1/2$, the effective Hamiltonian corresponding to the last four terms in Eq.~\ref{eq:Htotal}, is 
\begin{equation}
H_{\rm eff} = - 2 A_J g_J \mu_{\rm B} \, \textbf{B} \cdot \Lambda \cdot \textbf{I} - \gamma_{\rm n} \, \textbf{B} \cdot \textbf{I} - g^2_J \mu^2_{\rm B} \, \textbf{B} \cdot \Lambda \cdot \textbf{B} \, ,
\label{eq:Heff}
\end{equation}
where $A_J$ is the magnetic hyperfine constant, $g_J$ is the Land\'e factor of the electron, $\mu_{\rm B}$ is the Bohr magneton, $\textbf{B}$ the applied magnetic field, $\textbf{I}$ the nuclear spin, and $\gamma_{\rm n}$ the nuclear gyromagnetic ratio. The tensor $\Lambda$, expressed in second-order perturbation theory is
\begin{equation}
\Lambda_{\alpha \beta} = \sum_{n \neq 0} \frac{\left\langle 0 | J_{\alpha} | n \right\rangle \left\langle n | J_{\beta} | 0 \right\rangle}{E_n - E_0} \, .
\label{eq:Lambda}
\end{equation}
The sum is performed over the crystal field levels, which have energy $E_n$.

The first term of Eq.~\ref{eq:Heff} describes the second order coupling between the hyperfine interaction and the electronic Zeeman effect. Together with the second term, the nuclear Zeeman interaction, they form the so-called enhanced nuclear Zeeman interaction, which can be re-written as \cite{guillot-noel_analysis_2005}
\begin{equation}
\begin{aligned}
H_{\rm enZ} = -  \hbar \left( \gamma_{J,x} B_x I_x + \gamma_{J,y} B_y I_y + \gamma_{J,z} B_z I_z \right) \\ {\rm with} \; \gamma_{J,\alpha} = \gamma_{\rm n} + \frac{2 A_J g_J \mu_{\rm B} \Lambda_{\alpha \alpha}}{\hbar} \, ,
\end{aligned}
\label{eq:HenZ}
\end{equation}
where $\alpha = x, y$ or $ z$. In case of the $D_2$ symmetry of Tm:YAG, the $\Lambda$-tensor is diagonal in each local frame $(\textbf{x}, \textbf{y}, \textbf{z})$. This interaction leads to the splitting of the crystal field levels into spin states that are involved in persistent hole burning \cite{Ohlsson2003}. The effective gyromagnetic tensors $\gamma_J$ have been estimated theoretically for the ground and excited states of the optical $^3{\rm H}_6 \rightarrow \, ^3{\rm H}_4$ transition \cite{guillot-noel_analysis_2005}, and their components have been measured experimentally \cite{de_seze_experimental_2006,zafarullah_thulium_2008}. These $\gamma_J$ components are approximately 10 times larger in the ground state than in the excited state, which leads to much larger hyperfine splittings in the ground state.

\begin{figure}[t]
\centering
\includegraphics[width=0.5\columnwidth]{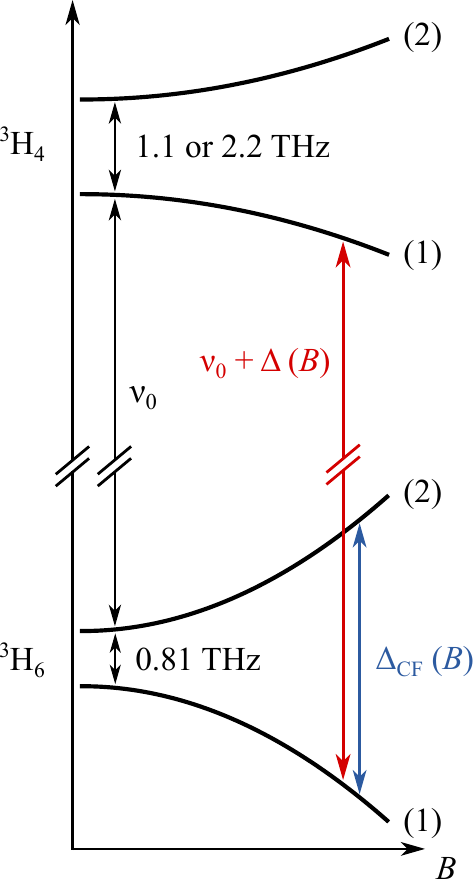}
\caption{Simplified energy level structure of Tm$^{3+}$:YAG, relevant for our study. Represented are the two lowest crystal field levels of the $^3$H$_6$ and $^3$H$_4$ multiplets \cite{gruber_spectra_1989,tiseanu_energy_1995}. In an external magnetic field, the crystal field level energies vary as $B^2$ due to the quadratic Zeeman effect, with different coefficients for each level. As a consequence, the optical transition $^3{\rm H}_6 (1) \rightarrow \, ^3{\rm H}_4 (1)$, which has a zero-field frequency of $\nu_0$, is shifted by $\Delta (B)$. The frequency $\Delta_{\rm CF}$, corresponding to the transition between the two crystal field levels in the ground state, also varies as $B^2$.}
\label{fig:levels_shift}
\end{figure}

The last term of Eq.~\ref{eq:Heff} corresponds to the second-order, or quadratic, electronic Zeeman interaction that results from the magnetic-field-induced mixing of the electronic crystal field wavefunctions. As the third crystal field level of the ground state is far from the lowest two levels, with an energy of 216 cm$^{-1}$, its contribution is negligible, so that this effect can be viewed as simple mixing between the lowest two crystal field levels. Because of this interaction, the applied magnetic field induces a weak magnetic moment, called Van Vleck paramagnetism, \cite{vleck_theory_1932} that shifts the energy of each crystal field level in both the ground and excited states, as illustrated in Fig.~\ref{fig:levels_shift}. This then leads to a frequency shift of the optical transition that is proportional to $B^2$ in the regime where the shifts are much smaller than the crystal field splittings.

Because a large enhancement of the effective nuclear moment arises from the coupling of the nuclear spin to the induced electronic paramagnetism \cite{abragam_enhanced_1983}, the quadratic Zeeman interaction can be directly related to the resulting hyperfine splittings \cite{teplov_magnetic_1968}. Thus, the energy displacement can be written in terms of the effective gyromagnetic tensor of the enhanced nuclear Zeeman interaction $\gamma_J$
\begin{equation}
\begin{aligned}
D_J =  \frac{g_J \mu_{\rm B} }{2 A_J }  & \left[  \left(\gamma_{J,x} - \gamma_{\rm n} \right) B_x^2 \right.\\ 
& \left. + \left(\gamma_{J,y} - \gamma_{\rm n} \right) B_y^2 + \left(\gamma_{J,z} - \gamma_{\rm n} \right) B_z^2 \right]  \, .
\end{aligned}
\label{eq:displacement}
\end{equation}
Finally, the frequency shift on the optical transition is given by $\Delta = (D_g - D_e) / h$ where $g $ and $e$ denote the ground and excited state, respectively. Because of the relation between excited and ground state hyperfine splittings and quadratic Zeeman effects, the observed hyperfine splittings can be used to infer the relative magnitude of the ground and excited state quadratic Zeeman effects. Moreover, since the components of $\gamma_J$ are much larger in the ground state than in the excited state, the quadratic Zeeman effect in the ground state is the largest contribution to the shift $\Delta$ of the optical transition. With this approximation, one should observe $\Delta (B) \sim \Delta_{\rm CF} (B)/2$, with $\Delta_{\rm CF}(B)$ the crystal field transition frequency (see Fig.~\ref{fig:levels_shift}).

\subsection{Measurements of the inhomogeneous line}

All measurements were performed on Tm$^{3+}$:YAG bulk crystals from Scientific Materials, with doping concentrations of either 0.1 or 1 \%. The sample was cooled  in an Oxford Instruments Spectromag cryostat through helium exchange gas. Our setup provided magnetic-field strengths of up to 6 T and sample temperatures $T$ down to 1.6 K.

\subsubsection{Magnetic field strength dependence}

We first studied the properties of the inhomogeneously broadened $^3{\rm H}_6 \rightarrow \, ^3{\rm H}_4$ absorption line (see Fig \ref{fig:levels_shift}) as we varied the magnetic field strength for a specific orientation of the magnetic field relative to the crystal. We initially chose to apply the magnetic field $\textbf{B}$ along the $[111]$ crystallographic axis (see Fig.~\ref{fig:yag}). In this case, the sites 1, 3, and 5 are magnetically equivalent, and the sites 2, 4, and 6 are magnetically equivalent. The light probing the medium with the $\textbf{k}$ vector along $[1\bar{1}0]$ was linearly polarized with the electric field vector $\textbf{E}$ parallel to the $[111]$ axis. In this case, the electric transition dipoles $\mu$ of site 2, 4, and 6 ions are orthogonal to the light polarization and thus only ions of sites 1, 3, and 5 are addressed (all featuring identical Rabi frequencies).

\begin{figure}[t]
\centering
\includegraphics[width=0.95\columnwidth]{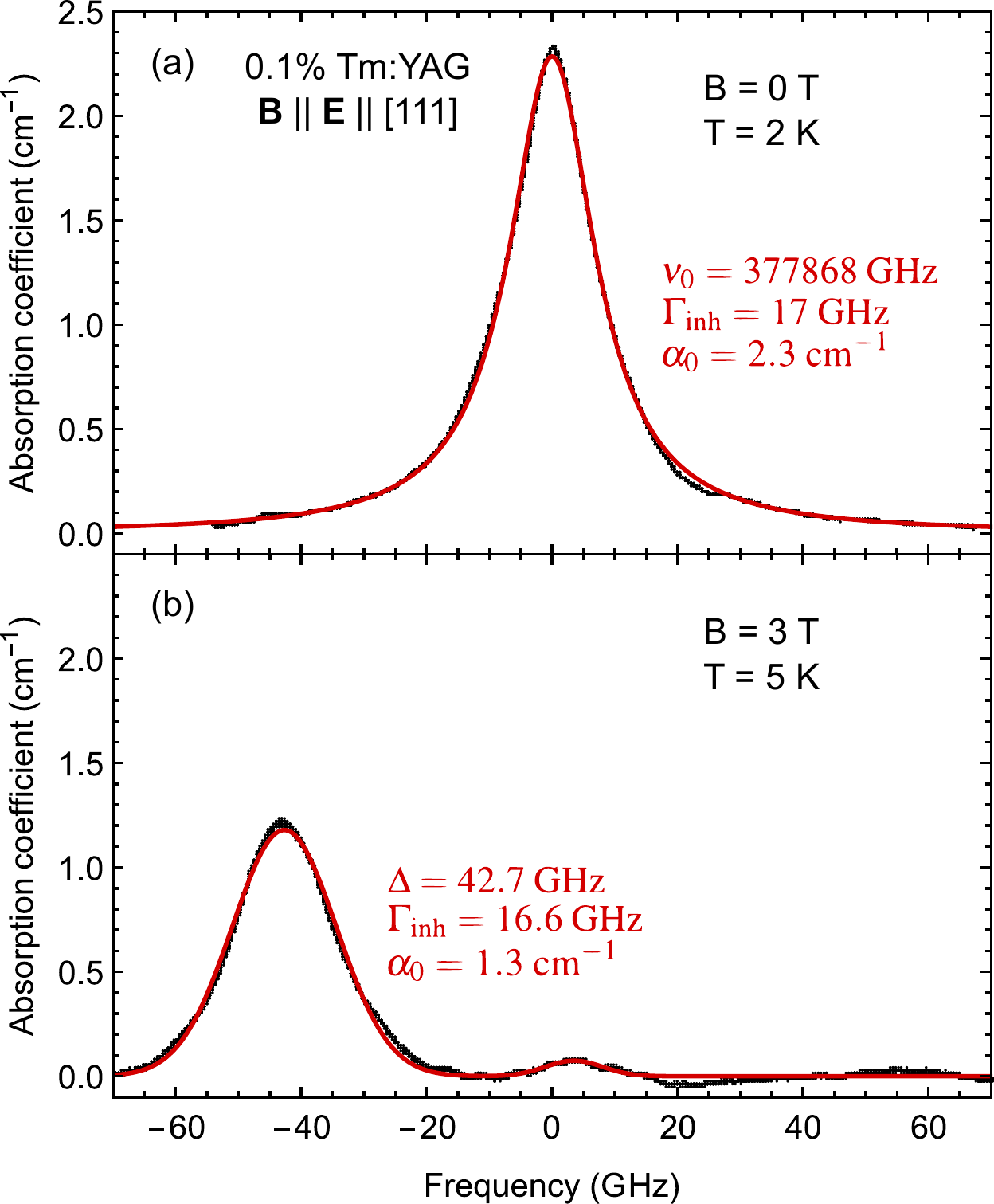}
\caption{Inhomogeneous line of a 0.1~\% Tm$^{3+}$:YAG crystal (a) at zero magnetic field and $T=2$ K, and (b) with a 3 T magnetic field along $[111]$ and at $T=5$ K. A narrowband laser probed the medium with $\textbf{k} \parallel [1\bar{1}0]$ and $\textbf{E} \parallel [111]$ (see main text for definitions). The experimental points were fit by a Lorientzian function with a full width at half maximum $\Gamma_{\rm inh} = 17$ GHz at zero magnetic field. The center frequency is $\nu_0 = 377 868$ GHz, and the absorption coefficient is $\alpha_0 = 2.3$ cm$^{-1}$. For $B=3$ T, the shift $\Delta$ is 42.7 GHz.}
\label{fig:inh_line}
\end{figure}

High resolution measurements of the inhomogeneous line were performed by scanning a New Focus Vortex laser with $\sim 1$ MHz linewidth over the $^3{\rm H}_6 \rightarrow \, ^3{\rm H}_4$ transition. The scans were calibrated via a Fabry Perot cavity with a free spectral range of 97 MHz. At zero magnetic field, the inhomogeneous line, shown in Fig.~\ref{fig:inh_line} (a), exhibits a Lorentzian shape with a full width at half maximum of 17 GHz, a center frequency of $\nu_0 = 377 868$ GHz (793.3788 nm) and a peak absorption coefficient of 2.3 cm$^{-1}$. When we applied a magnetic field along the $[111]$ axis, the inhomogeneous line was shifted by $\Delta$, as shown in Fig.~\ref{fig:inh_line} (b) where $B=3$ T and $\Delta = 42.7$ GHz.

\begin{figure}[t]
\centering
\includegraphics[width=0.75\columnwidth]{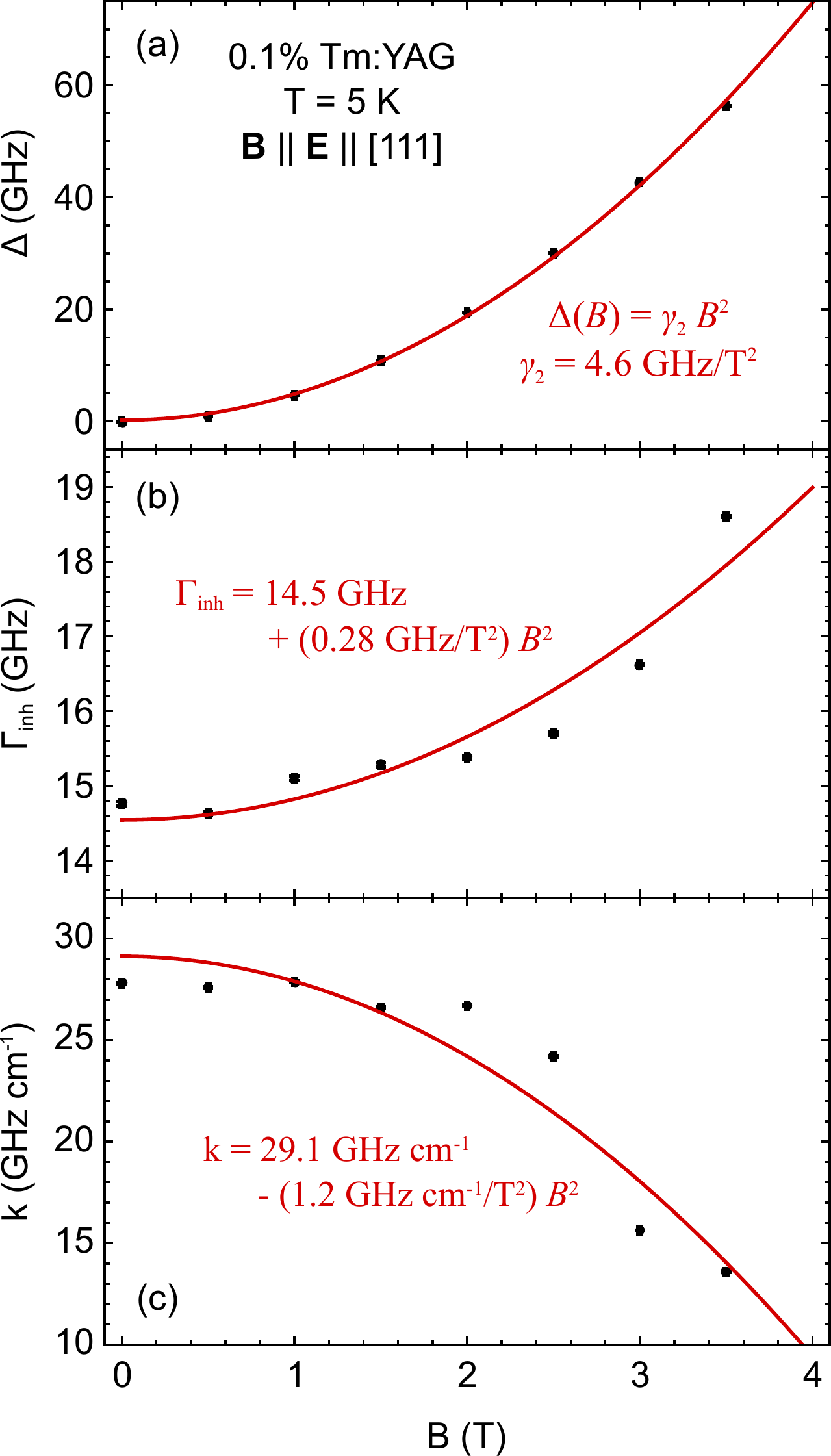}
\caption{Magnetic field dependence of several inhomogeneous line properties of a 0.1 \% Tm$^{3+}$:YAG crystal at $T=5$ K and with $\textbf{B} \parallel \textbf{E} \parallel [111] $. (a) Frequency shift $\Delta$ as a function of the magnetic field strength $B$. The experimental points were fit by a quadratic function, giving a coefficient $\gamma_2=4.69 \pm 0.03$ GHz/T$^2$. (b) Observed nonlinear increase in the inhomogeneous linewidth $\Gamma_{\rm inh}$ as a function of the magnetic field strength $B$. The experimental points were fit by a $B^2$ dependence, giving a broadening coefficient of 0.28 GHz/T$^2$. (c) Observed nonlinear decrease in the integrated transition linestrength as a function of the magnetic field strength $B$. The experimental points were fit by a $B^2$ dependence, giving a coefficient of -1.3 GHz cm$^{-1}$ T$^{-2}$.}
\label{fig:shift_B-dep}
\end{figure}

We then quantified the shift of the inhomogeneous line center frequency as the magnetic field strength $B$ was varied, as shown in Fig.~\ref{fig:shift_B-dep}. The experimental data were fitted by the quadratic function $\Delta (B)  = \gamma_2B^2$, yielding a coefficient $\gamma_2 = 4.69 \pm 0.03$ GHz/T$^2$. From Eq.~\ref{eq:displacement}, the energy displacements of the ground and excited states were calculated, using the known values of $A_J$ for $^{169}$Tm$^{3+}$ \cite{guillot-noel_analysis_2005} and the experimentally measured components of $\gamma_J$ \cite{de_seze_experimental_2006}. This lead to an expected value of 4.2 GHz/T$^2$, which is in agreement with the experimental one, supporting the validity of our model.

In addition to a shift in the transition energy due to the Zeeman effect, we also observed a nonlinear increase in the inhomogeneous linewidth $\Gamma_{\rm inh}$ and a decrease in the integrated absorption linestrength $k$ at higher fields. A quadratic increase in $\Gamma_{\rm inh}$ is expected due to the inhomogeneous broadening at zero field and the resulting spread of energy splittings in Eq.~\ref{eq:Lambda}. It is straightforward to show that if $\Gamma_{\rm CF}$ is the inhomogeneous broadening of the energy level splitting $\Delta_{\rm CF}$ between the two crystal field levels mixed by the quadratic Zeeman effect with a magnitude of $\gamma_{\rm CF}$, then the contribution to the optical inhomogeneous broadening will simply be
\begin{equation}
\Delta \Gamma_{\rm inh} =  \frac{ \gamma_{\rm CF} \Gamma_{\rm CF}}{ \Delta_{\rm CF} }  B^2  \, .
\label{eq:Ginh}
\end{equation}
Using our values of $\gamma_{\rm CF}$ and $\Delta_{\rm CF}$, obtained from measurements described in subsequent sections, we found that the observed coefficient of 0.28 GHz/T$^2$ would result entirely from the quadratic Zeeman effect assuming an inhomogeneous broadening of the crystal field energy level splitting of 27 GHz, which is within the range expected for this parameter, suggesting that this mechanism could explain the broadening.

In addition to a shift and broadening, a significant decrease in the integrated linestrength of the absorption was observed at higher fields. This effect likely results from a change in transition strength caused by the mixing of crystal field wavefunctions through the quadratic Zeeman effect. Consequently, we fit the measurements to a $B^2$ dependence, giving a coefficient of -1.3~GHz~cm$^{-1}$~T$^{-2}$.

\subsubsection{Orientation dependence}

Transmission spectra  were obtained by probing the optical $^3{\rm H}_6 \rightarrow \, ^3{\rm H}_4$ transition at 793 nm with broadband light from a tungsten-halogen lamp, and analyzing the transmitted light with an Advantest Q8347 optical spectrum analyzer with 900 MHz frequency resolution. A constant magnetic field of 6 T was applied, and the crystal was rotated to vary the orientation of the magnetic field by $90^{\circ}$ within a plane defined by the $[111]$ and $[\bar{1}\bar{1}2]$ crystal axes; the axis $[001]$ is also contained in the plane (see Fig.~\ref{fig:yag}). The angle $\theta$ denotes the angle between  $\textbf{B}$ and the $[111]$ axis. The light propagated along $[1\bar{1}0]$, with polarization either along $[111]$ or $[\bar{1}\bar{1}2]$. With $\textbf{E} \parallel [111] $, the electric transition dipoles $\mu$ of site 2, 4, and 6 ions are orthogonal to the light polarization and thus only ions of sites 1, 3, and 5 were addressed with identical Rabi frequencies. With $\textbf{E} \parallel [\bar{1}\bar{1}2]$, the electric transition dipole of site 2 is orthogonal to the light polarization, and the ratio $\vec{\mu} \cdot \textbf{E} / (\mu E)$, which is a measure of the interaction strength, is $\sqrt{3}/2 \approx 0.87$ for sites 4 and 6, $1/ \sqrt{3} \approx 0.58$ for site 1, and $1/(2 \sqrt{3}) \approx 0.29$ for sites 3 and 5.

\begin{figure}[t]
\centering
\includegraphics[width=.95\columnwidth]{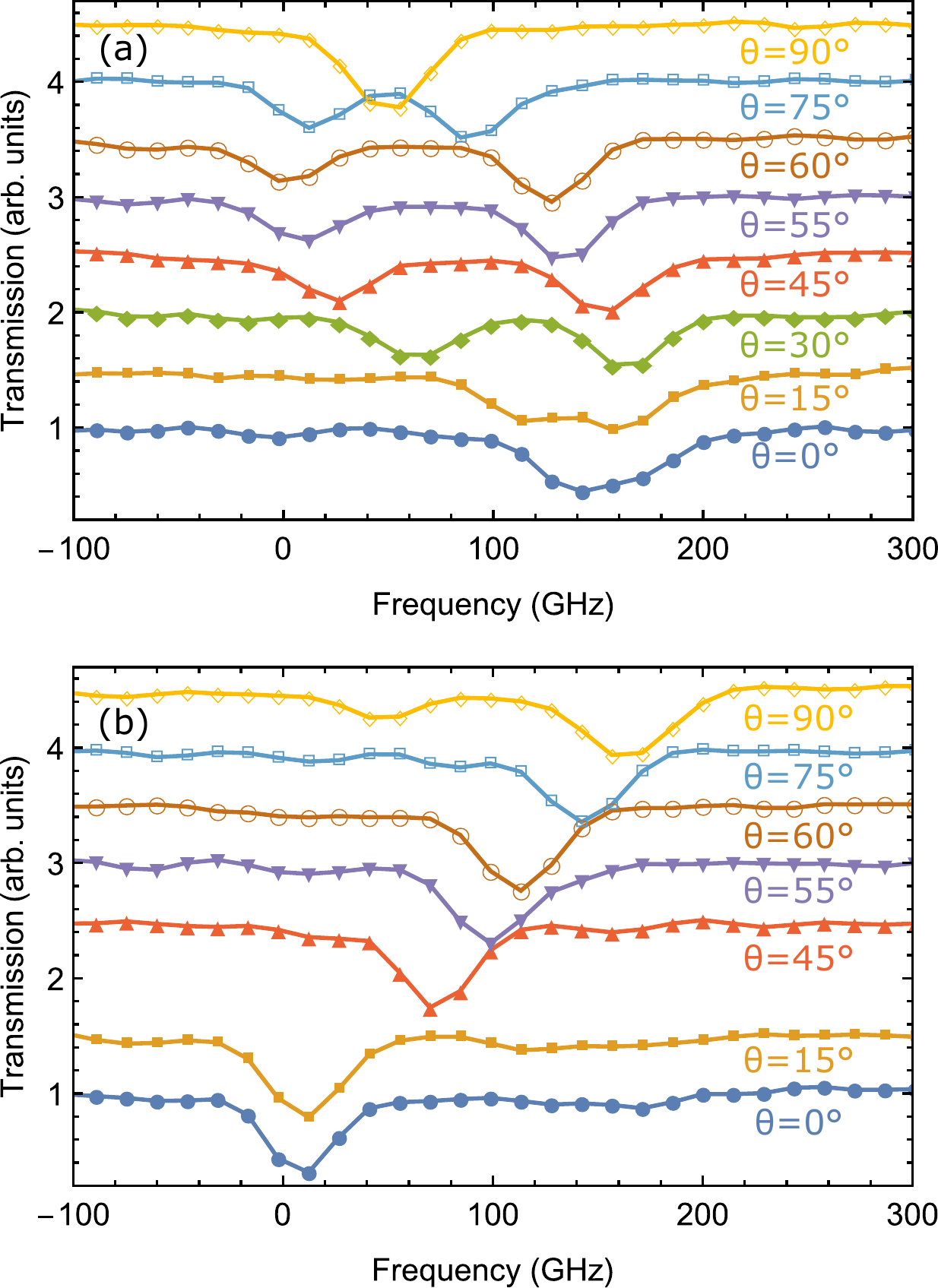}
\caption{Transmission spectra for various angles $\theta$ of magnetic field with respect to the $[111]$ crystal axis in the plane defined by the $[111]$ and $[\bar{1}\bar{1}2]$, of a 1\% Tm$^{3+}$:YAG crystal at 1.6 K. The strength of the applied magnetic field was 6 T. The polarization of the probing light was (a) along $[111]$, leading to interaction with ions of sites 1, 3, and 5, and (b) along $[\bar{1}\bar{1}2]$ leading to interaction with ions of sites 4 and 6.}
\label{fig:spectra}
\end{figure}

Fig.~\ref{fig:spectra} shows the obtained spectra, where zero frequency detuning was defined as $\nu_0 = 377 868$ GHz. The individual spectra were fit to extract the frequency shift $\Delta$. For panel (a), the spectra were recorded with the probe light polarized along $[111]$,  exciting only the ions of sites 1, 3, and 5. These three sites are magnetically equivalent for $\textbf{B} \parallel [111] $, and the first spectrum shows that they experience a large shift of about 150 GHz for $\theta =0$. As the magnetic field moved away from the $[111]$ axis, two different lines were observed, one corresponding to site 3- and 5-, the other to site 1-ions. For panel (b), the spectra were recorded with the polarization of the probe light along $[\bar{1}\bar{1}2]$, i.e. addressing mainly the ions of sites 4 and 6. The ions of those two sites are magnetically equivalent as long as the magnetic field is applied in the plane defined by the $[111]$ and $[\bar{1}\bar{1}2]$. Thus, they experience the same frequency shift, which is zero for $\textbf{B} \parallel [111] $ and reaches its maximum of around 160 GHz for $\textbf{B} \parallel [\bar{1}\bar{1}2]$. With this polarization, the interaction with sites 1, 3, and 5 is significantly reduced so that their absorption lines are only weakly visible in the spectra.

\begin{figure}[t]
\centering
\includegraphics[width=.95\columnwidth]{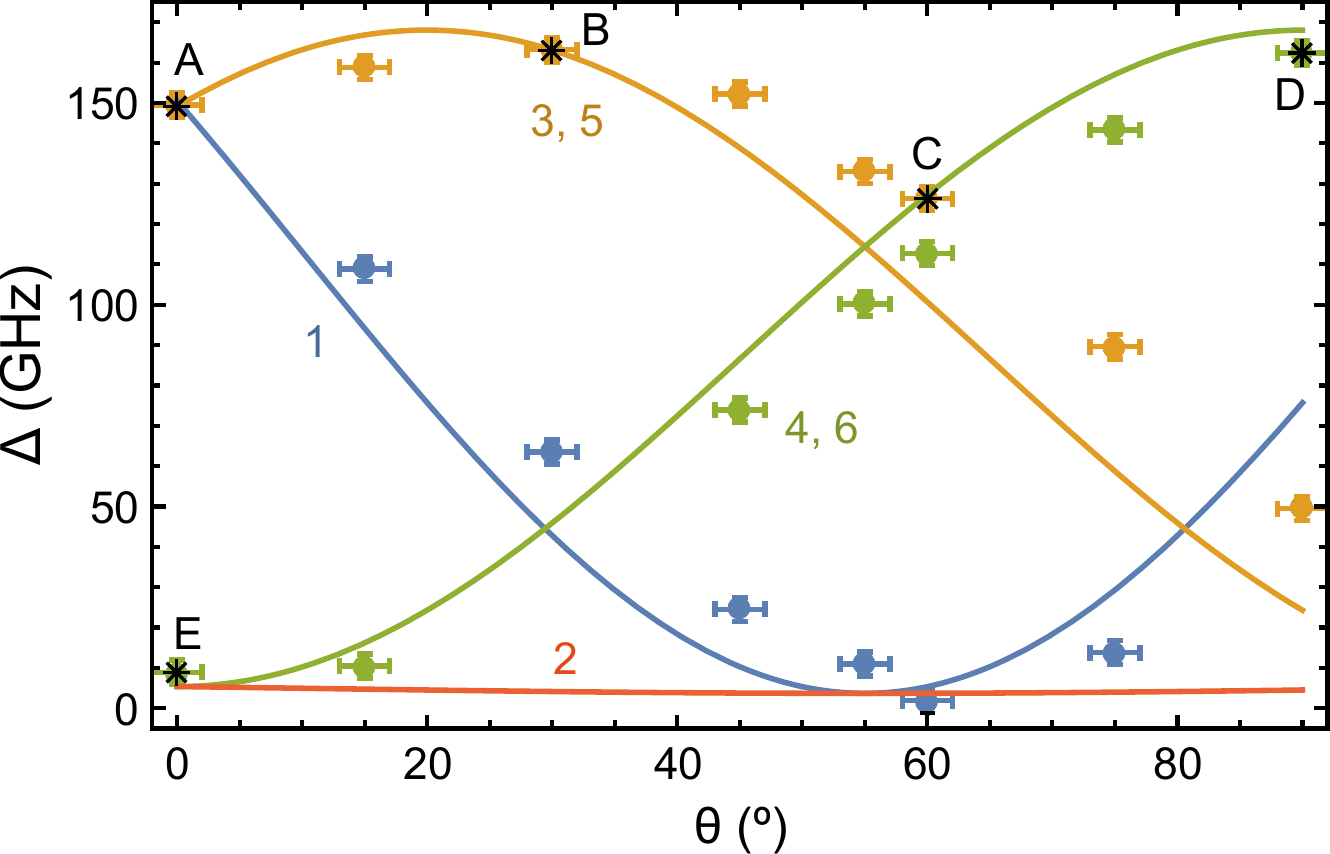}
\caption{Frequency shift $\Delta$ as a function of the magnetic field angle $\theta$ for ions of the site 1 (blue), sites 3 and 5 (yellow), and sites 4 and 6 (green), extracted from the transmission spectra shown in Fig.~\ref{fig:spectra}. The experimental points are presented together with the expected values (solid lines). The points labeled by letters (A-E) identify configurations for which the spin-state lifetime was measured (see Table \ref{tab:T1_orient})}
\label{fig:shiftorientdep}
\end{figure}

All frequency shifts obtained for the different configurations are shown in Fig.~\ref{fig:shiftorientdep} as a function of the angle of the magnetic field $\theta$. Due to the limited precision of the mechanical rotation mount used to hold the crystal, the uncertainty in the magnetic field angle was $\pm\ 2^{\circ}$ . For each site, the energy displacements of the ground and excited states were determined from Eq.~\ref{eq:displacement} using the known values of $A_J$ for $^{169}$Tm$^{3+}$ \cite{guillot-noel_analysis_2005} and the experimentally measured components of $\gamma_J$ \cite{de_seze_experimental_2006}. The theoretical orientation dependence of the shift is also plotted in Fig.~\ref{fig:shiftorientdep}.  Experimental and semi-theoretical values are in relatively good agreement.

Our study revealed that at magnetic field strengths of a few Tesla and for certain orientations of the field, site-selectivity by tuning the laser frequency to follow the corresponding quadratic Zeeman shifts is possible e.g. by tuning the laser by 160 GHz from $\nu_0$ mainly ions at sites 4 and 6 will be selected at a field of 6 Tesla. Due to its independence of polarization, this method to select different ions is beneficial if one has random polarization, for example when working with powders that scramble polarization through scattering \cite{lutz_effects_2016}.

\section{Nuclear spin-state lifetimes}
The above characterization of the quadratic Zeeman shift was employed to study the lifetimes of the spin states in a high magnetic field. Using persistent spectral hole burning methods, we monitored the relaxation of optically pumped non-equilibrium spin populations \cite{macfarlane_coherent_1987}. Light from a Coherent 899-21 Ti:Sapphire laser was sent to the sample for 100 ms to burn a spectral hole. After a variable wait time, the laser was scanned across 180 MHz in 250 $\mu$s via a double-passed 200 MHz acousto-optic modulator to probe the spectral hole. The power of the laser beam focused into the crystal was $\sim 10$ $\mu$W. After the sample, the transmitted pulse was measured with a New Focus 2051 detector. The decay of the spectral hole area as a function of the waiting time enables one to obtain the lifetime of the spin states, which can be as long as hours for Tm$^{3+}$:YAG at 1.6 K.

\subsection{Orientation dependence}

The spin-state lifetime $T_1$ at $B= 6$ T and $T=1.6$ K was first measured in a 1\% Tm:YAG crystal for different orientations of the magnetic field and light polarization relative to the crystal. The frequency of the laser was tuned to follow the inhomogeneous line of the investigated transitions, according to each line's different quadratic Zeeman shift. The polarization direction of the light was kept either along the magnetic field or perpendicular to it. Table \ref{tab:T1_orient} shows the spin-state lifetimes $T_1$ for the different configurations of laser frequencies, magnetic field orientations, and polarizations.

We observed that the spin-state lifetime significantly changes as a function of the magnetic field orientation and depends on which subset of sites are addressed. The longest $T_1$ were found for the subset of ions at sites 1, 3, and 5, in the configuration in which they are magnetically and optically equivalent and experience a large shift of the optical transition due to the quadratic Zeeman effect, i.e. $\textbf{B} \parallel \textbf{E} \parallel [111]$. In this case the spin-state lifetime is $\sim 1$ hour, whereas it is only $\sim 6$ minutes for site 2 with the same magnetic field strength and orientation. This suggests that different processes dominate the spin-relaxation depending on the site. In the E configuration (see Table \ref{tab:T1_orient}) , the projection of the $g$-tensor of site 2 on the magnetic field direction is small \cite{de_seze_experimental_2006}. In this case, the process responsible for spin-relaxation could be mutual spin flips caused by the magnetic dipole-dipole interaction between Tm$^{3+}$ ions and other nuclear spins in the lattice \cite{ahlefeldt_optical_2015}, or relaxation induced by trace paramagnetic impurities in the crystal \cite{abragam_principles_1961,bleaney_nuclear_1983,aminov_nuclear_1985}. In the following, we focused on the subset of ions at sites 1, 3, and 5 in order to understand the spin-relaxation mechanisms that limit the longest nuclear spin-state lifetimes.

\begin{table}[t]
  \centering
  \begin{tabular}{|c|c|c|c|c|}
  \hline
   label & sites & $\textbf{B}$ and $\textbf{E}$ & shift & $T_1$ \\
   \hline
   A & 1, 3, 5 & $\textbf{B} \parallel \textbf{E} \parallel [111]$ & 153 GHz & 65 $\pm$ 4 min. \\
   \hline
  B &  3, 5 & $30 ^{\circ}$ from $[111]$ & 160 GHz & 50 $\pm$ 2 min. \\
    \hline
  C &  3, 5 & $\textbf{B} \parallel \textbf{E} \parallel [001]$ & 120 GHz & 18 $\pm$ 1 min. \\
    \hline
  D &  4, 6 & $\textbf{B} \parallel \textbf{E} \parallel [\bar{1}\bar{1}2]$ & 165 GHz & 38 $\pm$ 8 min. \\
   \hline
    E &  2 & $\textbf{B} \parallel [111]$, $ \textbf{E} \parallel [\bar{1}\bar{1}2]$ & 0 GHz & 5.9 $\pm$ 1.7 min. \\
   \hline
  \end{tabular}
  \caption{Nuclear spin-state lifetimes for ions at different sites in a magnetic field of 6 T and at 1.6 K in  1 \% Tm$^{3+}$:YAG. Each measurement is labeled and displayed in Fig.~\ref{fig:shiftorientdep}.}
  \label{tab:T1_orient}
\end{table}

\subsection{Spin-lattice relaxation}

Next, the nuclear spin-state lifetime was studied for ions at sites 1, 3, and 5, which offer the longest relaxation time (see above). We used the configuration in which they are optically and magnetically equivalent, i.e. with the magnetic field and the light polarization along $[111]$. For this experiment we chose a lower concentration 0.1\% Tm$^{3+}$:YAG crystal. The frequency of the laser was adjusted to follow the shift of the optical transition due to the quadratic Zeeman effect as $B$ was increased. We investigated the temperature and magnetic field dependences of the nuclear spin-state lifetime for 1.6 K $\leq T \leq 4.5$ K and 0 T $\leq B \leq 6$ T, as shown in Fig.~\ref{fig:T-dep_3T}, \ref{fig:B-dep_1.6K} and \ref{fig:B-dep_4K}. For this range of parameters, we identified two processes limiting the spin-state lifetime: the direct single-phonon process, and the two-phonon Orbach process. As we will describe below, the direct phonon process, which results from the absorption or emission of a phonon resonant with the hyperfine splitting in the ground state, is dominant at low temperature and high magnetic field. The Orbach process, in which two higher energy phonons induce relaxation by driving transitions to and from other electronic crystal field levels, is dominant at higher temperature.

\subsubsection{Temperature dependence}

\begin{figure}[t]
\centering
\includegraphics[width=.95\columnwidth]{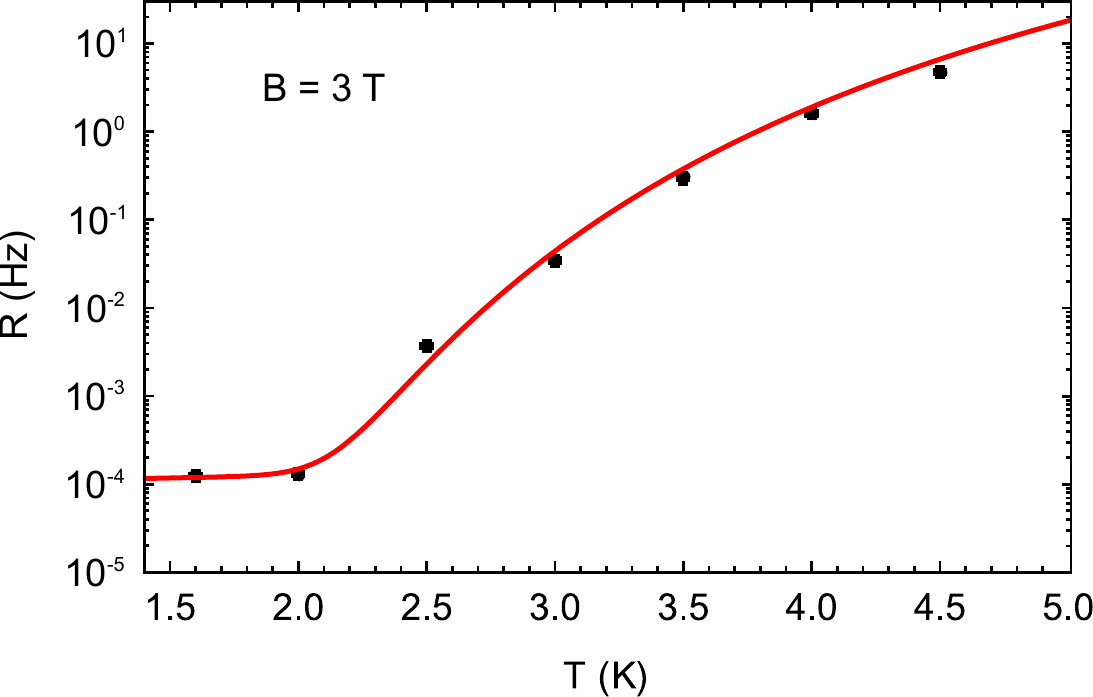}
\caption{Spin-lattice relaxation rate as a function of the temperature $T$ for $B=3$ T in 0.1\% Tm:YAG for ions of sites 1, 3, and 5. The magnetic field and the light polarization were along $[111]$, and ions of sites 1, 3, and 5 were thus optically and magnetically equivalent. The experimental points were fit by Eq.~\ref{eq:spin-relax-rate} with the parameters given in Table~\ref{tab:fitparameters}.}
\label{fig:T-dep_3T}
\end{figure}

Two regimes of relaxation are easily visible in Fig.~\ref{fig:T-dep_3T}, which depicts the temperature dependence of the spin-relaxation rate $R$, i.e. the inverse of the nuclear spin-state lifetime $T_1$, at a magnetic field of 3 T. For temperatures above 2 K, the relaxation rate increases very rapidly, which is characteristic for a relaxation mechanism involving two phonons. Indeed, the rate of the Orbach process varies as $\sim e^{- \Delta_{\rm CF} / k_{\rm B} T}$, where $\Delta_{\rm CF}$ is the transition frequency between the two lowest crystal field levels and $k_{\rm B}$ the Boltzmann constant \cite{orbach_spin-lattice_1961}. Below $T = 2$ K, the spin relaxation rate increases linearly with temperature, characteristic of the direct phonon process. However, the temperature dependence of the spin-relaxation rate is not sufficient to identify the direct phonon process since other mechanisms, such as the magnetic dipole-dipole interaction between rare-earth ions that leads to mutual spin flips, can also have a linear temperature dependence \cite{Bottger2006}. More generally, interactions with paramagnetic impurities in the host matrix can also give rise to a linear temperature dependence of the spin relaxation rate \cite{abragam_principles_1961,bleaney_nuclear_1983,aminov_nuclear_1985}. However, the magnetic field dependence of the nuclear spin-state lifetime allowed us to unambiguously identify the direct process, as detailed below.

\subsubsection{Magnetic field dependence at 1.6 K}

\begin{figure}[t]
\centering
\includegraphics[width=.9\columnwidth]{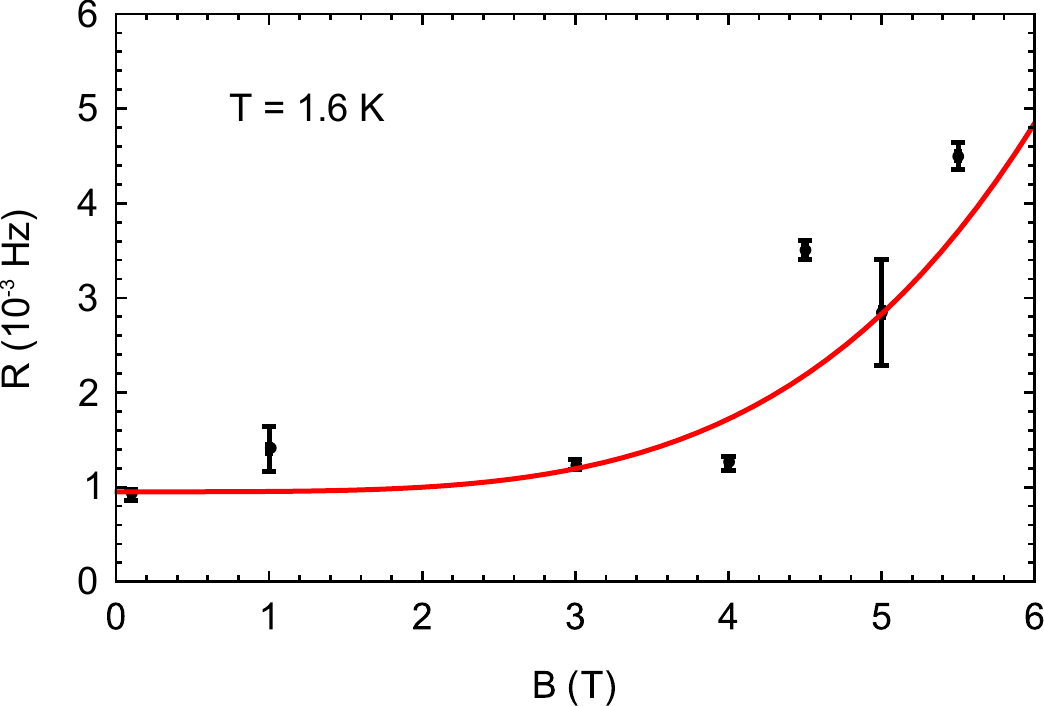}
\caption{Spin-lattice relaxation rate as a function of the magnetic field strength $B$ at $T=1.6$ K in 0.1\% Tm:YAG for ions of sites 1, 3, and 5. The magnetic field and the light polarization were along $[111]$. Ions of sites 1, 3, and 5 were optically and magnetically equivalent. The experimental points were fit by Eq.~\ref{eq:spin-relax-rate} with the parameters given in Table~\ref{tab:fitparameters}.}
\label{fig:B-dep_1.6K}
\end{figure}

To identify the direct phonon process, we studied the nuclear spin-state lifetime as a function of the magnetic field strength $B$ at low temperatures, i.e. below 2 K. When the thermal energy is much larger than the level splittings, as in the case considered here, the magnetic field dependence of the direct phonon process is $B^2$ for non-Kramers ions and $B^4$ for Kramers ions \cite{orbach_spin-lattice_1961}. Since the enhanced nuclear Zeeman splitting of the spin $1/2$ states of $^{169}$Tm$^{3+}$ arises from the induced electronic paramagnetism \cite{abragam_enhanced_1983}, the spin-lattice relaxation is thus expected to be proportional to $B^4$. Fig.~\ref{fig:B-dep_1.6K} shows the spin relaxation rate at $T=1.6$~K, which increases with the magnetic field strength as $B^4$. We can thus confirm that the direct phonon process dominates the spin relaxation of Tm:YAG at 1.6 K and magnetic field above 3 T.

\subsubsection{Magnetic field dependence at 4 K}

\begin{figure}[t]
\centering
\includegraphics[width=.95\columnwidth]{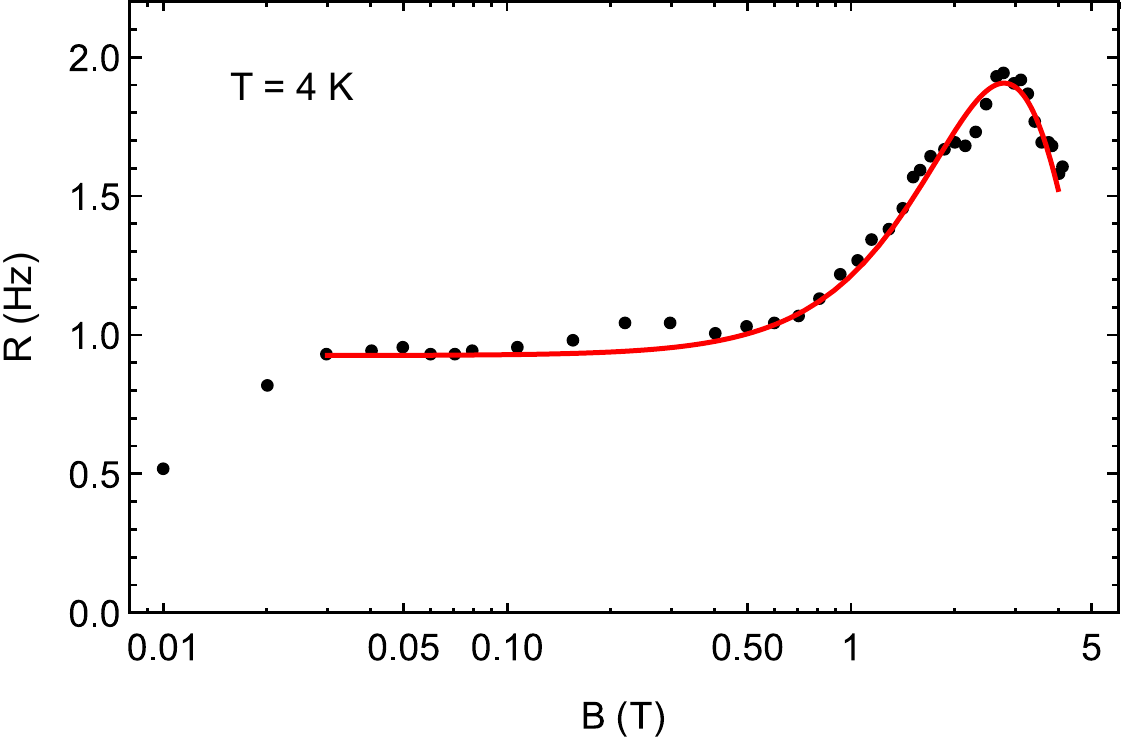}
\caption{Spin-lattice relaxation rate as a function of the magnetic field strength $B$ at $T=4$ K in 0.1\% Tm:YAG for ions of sites 1, 3, and 5. The magnetic field and the light polarization were along $[111]$, so ions of sites 1, 3, and 5 were optically and magnetically equivalent. The experimental points were fit by Eq.~\ref{eq:spin-relax-rate} with the parameters given in Table~\ref{tab:fitparameters}.}
\label{fig:B-dep_4K}
\end{figure}

We also investigated the spin relaxation rate at $T=4$~K, a temperature at which the Orbach process is dominant. The rate of this process is generally expected to be independent of magnetic field because it depends only on the energy spacing between the crystal field levels. The experimental points are shown in Fig.~\ref{fig:B-dep_4K}, and exhibit a slight increase between 1 and 3 T, followed by a decrease. As we describe below in more detail, this behavior can be explained by taking into account the increase in the energy splitting of the two lowest crystal field levels due to the quadratic Zeeman effect, as shown in Fig.~\ref{fig:levels_shift}. 

\subsubsection{Model}

Following the conclusions from the above discussion, we assume that for our range of parameters (1.6 K $\leq T \leq 4.5$ K and 0 T $\leq B \leq 6$ T), the spin relaxation rate can be written as
 \begin{equation}
R(B,T) = R_0 + \alpha_{\rm D}\,  \gamma^2 \,  B^4 \, T +  \frac{\alpha_{\rm O} (B)}{e^{\Delta_{\rm CF}(B)/ k_{\rm B} T }-1} \, ,
\label{eq:spin-relax-rate}
\end{equation}
with $\alpha_{\rm O} (B) = \alpha + \beta B^2$ and $\Delta_{\rm CF}(B)=\Delta_{\rm CF}^0 + \gamma_{\rm CF} B^2$. The first term corresponds to the residual relaxation at zero magnetic field and temperature. The second term describes the direct phonon process \cite{orbach_spin-lattice_1961}, with the coefficient $\alpha_{\rm D}$ depending on the properties of the host material such as the speed of sound and crystal density, and $\gamma$ represents the effective nuclear gyromagnetic ratio (in frequency units) along the magnetic field direction for the ground state hyperfine levels, equal to 400 MHz/T in this orientation \cite{de_seze_experimental_2006}. This term is valid for $\gamma B \ll k_{\rm B} T$, which is the case for our measurements. The last term corresponds to the Orbach process, with a coupling coefficient $\alpha_{\rm O}$ \cite{orbach_spin-lattice_1961}. To take into account the shift of the transition between the crystal field levels arising from the quadratic Zeeman effect, we include the corresponding $B^2$ dependence in the coupling coefficient $\alpha_{\rm O}$ and in the energy difference $\Delta_{\rm CF}$. 

\begin{table}[t]
  \centering
  \begin{tabular}{|c|c|}
  \hline
   $R_0$ & $9.5 \times 10^{-5}$ Hz \\
   \hline
   $\alpha_{\rm D}$ & $1.2 \times 10^{-24}$ (Hz K T$^2$)$^{-1}$ \\
   \hline
  $\alpha$ & $3.0 \times 10^4$ Hz \\
    \hline
  $\beta$ &  $1.3 \times 10^4$ Hz T$^{-2}$ \\
    \hline
  $\Delta_{\rm CF}^0$ & $8.3 \times 10^{11}$ Hz \\
   \hline
   $\gamma_{\rm CF}$ & $8.0 \times 10^{9}$ Hz T$^{-2}$ \\
   \hline
  \end{tabular}
  \caption{Parameters obtained from the 2-dimensional fit of the experimental spin relaxation rate data by Eq.~\ref{eq:spin-relax-rate}.}
  \label{tab:fitparameters}
\end{table}

We simultaneously fit the three experimental data sets shown in Figures \ref{fig:T-dep_3T}, \ref{fig:B-dep_1.6K} and \ref{fig:B-dep_4K} using a two dimensional fit, with the magnetic field and temperature as variables, to our model (\ref{eq:spin-relax-rate}) describing the relaxation. We obtained the set of parameters given in Table~\ref{tab:fitparameters}, which describes all of the observed behavior well. From this fit we find a direct phonon coefficient of $\alpha_D = (1.2 \pm 0.2) \times 10^{-24}$ (Hz K T$^2$)$^{-1}$. We compare this result to theoretical estimates using the method of Bleaney et al \cite{bleaney_nuclear_1983}. In their approach, the electron-phonon coupling matrix elements are approximated by order-of-magnitude estimates, giving the simple analytical relation
\begin{equation}
\alpha_{\rm D} \sim \frac{24 \pi^2 k_{\rm B} \gamma^2}{\rho v^5} \, ,
\label{eq:direct_coupling}
\end{equation}
where $\rho$ is the crystal density, and $v$ is the average acoustic phonon velocity.  For YAG, the density is $\rho = 4564$~kg/m$^3$ \cite{Stoddart1993} and the longitudinal and transverse acoustic velocities are $v_l = 8600$~m/s and $v_t = 5000$~m/s \cite{spencer_microwave_1963}. If we approximate $v$ by the acoustic velocity averaged over the single longitudinal mode and the two transverse modes, then, for our field orientation where $\gamma = 400$ MHz/T, Bleaney's approach predicts a direct phonon coefficient on the order of $10^{-26}$ (Hz K T$^2$)$^{-1}$, somewhat smaller than our observed value but within the range expected considering the approximations involved in the theoretical estimate.

The only cases that we are aware of in which direct phonon relaxation has been suggested for $^{169}$Tm$^{3+}$ have been related to studies of TmPO$_4$ and TmVO$_4$ materials for nuclear cooling applications \cite{suzuki_squid_1980,suzuki_enhanced_1981,roinel_relaxation_1985}. In these works, the observed relaxation rates require exceedingly large phonon coupling coefficients of $6 \times 10^{-11}$ (Hz K T$^2$)$^{-1}$ in TmPO$_4$ \cite{suzuki_enhanced_1981} and $4 \times 10^{-8}$ (Hz K T$^2$)$^{-1}$ in TmVO$_4$ \cite{suzuki_enhanced_1981}. It was subsequently proposed that the observed relaxation was due to paramagnetic impurities rather than phonon interactions \cite{bleaney_nuclear_1983,aminov_nuclear_1985}, although explanations in terms of phonon bottleneck effects were also proposed \cite{roinel_relaxation_1985}.  Considering the small value of $\alpha_{\rm D}$ observed in our work and the reasonable agreement of this value with theoretical predictions, our results support this conclusion. Thus, to the best of our knowledge, there have been no other unambiguous observations of direct phonon relaxation of the nuclear hyperfine states of $^{169}$Tm$^{3+}$ other than the ones presented here for Tm$^{3+}$:YAG.

Our fit of Eq.~\ref{eq:direct_coupling} results in a value of $0.83$~THz for the crystal field level transition frequency $\Delta_{\rm CF}$, in agreement with the value of 0.81 THz usually cited in the literature from direct spectroscopic measurements \cite{macfarlane_direct_2000}. Furthermore, we found $\gamma_{\rm CF} = 8.0$ GHz/T$^2$ for the coefficient of the quadratic Zeeman shift of the splitting between the first two crystal field levels in the $^3{\rm H}_6$ ground state. As discussed in the theoretical section, it is likely that the shift of the optical transition would be mainly due to the splitting between the crystal field levels in the ground state, and so that the coefficients would be approximated by $\gamma_{\rm CF} \sim 2 \gamma_2 \sim 9$ GHz/T$^2$. The value given by our fit is very close to this limit, supporting our analysis.


\section{Conclusion}

In summary, we characterized the shifts of the optical $^3{\rm H}_6 \rightarrow \, ^3{\rm H}_4$ transition in Tm$^{3+}$:YAG resulting from the quadratic Zeeman effect for the ground and excited states in large magnetic fields. The orientation dependence of the shifts is in agreement with the theoretical predictions. A coefficient of 4.69 GHz/T$^2$ is measured in the case where the sites 1, 3, and 5 are optically and magnetically equivalent. This shift provides a tool to spectrally select ions at different sites. 

We furthermore measured the spin lifetimes for different sites and orientations of the magnetic field, with a maximum value observed for sites 1, 3, and 5 when they are optically and magnetically equivalent. In that configuration, the spin lifetime was investigated as a function of magnetic field strength and temperature, enabling us to identify the processes responsible for spin relaxation, i.e. the direct phonon process at low temperature ($T \leq 2$ K) and high magnetic field ($B \geq 3$ T), and the Orbach process at higher temperature ($T \geq 2$ K) and fields up to 6 T. We also found that the small increase in the energy splitting of the electronic crystal field levels due to the quadratic Zeeman effect can result in a significant decrease in the two-phonon Orbach process rate at low temperatures, introducing a magnetic field dependence into the otherwise field-independent relaxation process. Consequently, it is essential to consider this additional effect when analyzing spin-lattice relaxation of this and other rare-earth ions at higher magnetic field strengths. This finding potentially also affects the interpretation of past spin relaxation studies at low temperatures.

\section{Acknowledgments}

The authors acknowledge support from Alberta Innovates Technology Futures (ATIF), the National Engineering and Research Council of Canada (NSERC), and the National Science Foundation of the USA (NSF) under award nos. PHY-1212462, PHY-1415628, and CHE-1416454. W. T. is a senior fellow of the Canadian Institute for Advance Research (CIFAR).


\begin{thebibliography}{10}
\newcommand{\enquote}[1]{``#1''}

\bibitem{cole_coherent_2002}
Z.~Cole, T.~B\"{o}ttger, R.~K. Mohan, R.~Reibel, W.~R. Babbitt, R.~L. Cone, and
  K.~D. Merkel, Applied Physics Letters \textbf{81}, 3525 (2002).

\bibitem{lavielle_wideband_2003}
V.~Lavielle, I.~Lorger\'{e}, J.-L. Le~Gou\"{e}t, S.~Tonda, and D.~Dolfi, Optics
  Letters \textbf{28}, 384 (2003).

\bibitem{strickland_laser_2000}
N.~M. Strickland, P.~B. Sellin, Y.~Sun, J.~L. Carlsten, and R.~L. Cone,
  Physical Review B \textbf{62}, 1473 (2000).

\bibitem{lauro_slow_2009}
R.~Lauro, T.~Chaneli\`{e}re, and J.~L. Le~Gou\"{e}t, Physical Review A
  \textbf{79}, 063844 (2009).

\bibitem{bonarota_efficiency_2010}
M.~Bonarota, J.~Ruggiero, J.~L. Le~Gou\"{e}t, and T.~Chaneli\`{e}re, Physical
  Review A \textbf{81}, 033803 (2010).

\bibitem{macfarlane_photon-echo_1993}
R.~M. Macfarlane, Optics Letters \textbf{18}, 1958 (1993).

\bibitem{thiel_measuring_2014}
C.~W. Thiel, R.~M. Macfarlane, Y.~Sun, T.~Böttger, N.~Sinclair, W.~Tittel, and
  R.~L. Cone, Laser Physics \textbf{24}, 106002 (2014).

\bibitem{abragam_enhanced_1983}
A.~Abragam and B.~Bleaney, Proceedings of the Royal Society of London. Series
  A, Mathematical and Physical Sciences \textbf{387}, 221 (1983).

\bibitem{macfarlane_spectral_1987}
R.~M. Macfarlane and J.~C. Vial, Physical Review B \textbf{36}, 3511 (1987).

\bibitem{ohlsson_long-time-storage_2003}
N.~Ohlsson, M.~Nilsson, S.~Kr\"{o}ll, and R.~K. Mohan, Optics Letters
  \textbf{28}, 450 (2003).

\bibitem{goldner_hyperfine_2006}
P.~Goldner, O.~Guillot-Noël, A.~Louchet, F.~de~Sèze, V.~Crozatier,
  I.~Lorgeré, F.~Bretenaker, and J.~L. Le~Gouët, Optical Materials
  \textbf{28}, 649 (2006).

\bibitem{louchet_branching_2007}
A.~Louchet, J.~S. Habib, V.~Crozatier, I.~Lorger\'{e}, F.~Goldfarb,
  F.~Bretenaker, J.-L.~Le Gou\"{e}t, O.~Guillot-No\"{e}l, and P.~Goldner,
  Physical Review B \textbf{75}, 035131 (2007).

\bibitem{sun_symmetry_2000}
Y.~Sun, G.~M. Wang, R.~L. Cone, R.~W. Equall, and M.~J.~M. Leask, Physical
  Review B \textbf{62}, 15443 (2000).

\bibitem{dillon_ferrimagnetic_1961}
J.~F. Dillon and L.~R. Walker, Physical Review \textbf{124}, 1401 (1961).

\bibitem{tongning_ralentir_2014}
R.-C. Tongning, \enquote{Ralentir le déphasage des états de superposition
  atomiques dans un cristal de {Tm}3+:{YAG}}, Ph.D. thesis, Université Paris
  Sud - Paris XI (2014).

\bibitem{suzuki_squid_1980}
H.~Suzuki, Y.~Higashino, and T.~Ohtsuka, Journal of Low Temperature Physics
  \textbf{41}, 449 (1980).

\bibitem{suzuki_enhanced_1981}
H.~Suzuki, T.~Inoue, and T.~Ohtsuka, North-Holland Publishing Company
  \textbf{Physics}, 563 (1981).

\bibitem{roinel_relaxation_1985}
Y.~Roinel, V.~Bouffard, J.-F. Jacquinot, C.~Fermon, and G.~Fournier, Journal de
  Physique \textbf{46}, 1699 (1985).

\bibitem{macfarlane_coherent_1987}
R.~M. Macfarlane and R.~M. Shelby, in \emph{Modern {Problems} in {Condensed}
  {Matter} {Sciences}}, vol.~21 of \emph{Spectroscopy of {Solids} {Containing}
  {Rare} {Earth} {Ions}}, A.~A. Kaplyanskii and R.~M. Macfarlane, eds.
  (Elsevier, 1987), pp. 51--184.

\bibitem{teplov_magnetic_1968}
M.~A. Teplov, Journal of Experimental and Theoretical PhysicsSoviet Physics
  Jetp \textbf{26}, 872 (1968).

\bibitem{guillot-noel_analysis_2005}
O.~Guillot-No\"{e}l, P.~Goldner, E.~Antic-Fidancev, and J.~L. Le~Gou\"{e}t,
  Physical Review B \textbf{71}, 174409 (2005).

\bibitem{Ohlsson2003}
N.~Ohlsson, M.~Nilsson, S.~Krll, and R.~K. Mohan, Optics Letters \textbf{28},
  450 (2003).

\bibitem{de_seze_experimental_2006}
F.~de~Seze, A.~Louchet, V.~Crozatier, I.~Lorger\'{e}, F.~Bretenaker, J.-L.
  Le~Gou\"{e}t, O.~Guillot-No\"{e}l, and P.~Goldner, Physical Review B
  \textbf{73}, 085112 (2006).

\bibitem{zafarullah_thulium_2008}
I.~Zafarullah, \enquote{Thulium {Ions} in a {Yttrium} {Aluminum} {Garnet}
  {Host} for {Quantum} {Computing} {Applications}: {Material} {Analysis} and
  {Single} {Qubit} {Operations}}, Ph.D. thesis, Montana State University,
  Bozeman, Montana (2008).

\bibitem{gruber_spectra_1989}
J.~B. Gruber, M.~E. Hills, R.~M. Macfarlane, C.~A. Morrison, G.~A. Turner,
  G.~J. Quarles, G.~J. Kintz, and L.~Esterowitz, Physical Review B \textbf{40},
  9464 (1989).

\bibitem{tiseanu_energy_1995}
C.~Tiseanu, A.~Lupei, and V.~Lupei, Journal of Physics: Condensed Matter
  \textbf{7}, 8477 (1995).

\bibitem{vleck_theory_1932}
J.~H.~V. Vleck, \emph{The theory of electric and magnetic susceptibilities}
  (Oxford University Press, 1932).

\bibitem{lutz_effects_2016}
T.~Lutz, L.~Veissier, C.~W. Thiel, P.~J.~T. Woodburn, R.~L. Cone, P.~E.
  Barclay, and W.~Tittel, Science and Technology of Advanced Materials
  \textbf{17}, 63 (2016).

\bibitem{ahlefeldt_optical_2015}
R.~L. Ahlefeldt, M.~F. Pascual-Winter, A.~Louchet-Chauvet, T.~Chaneli\`{e}re, and
  J.-L. Le~Gou\"{e}t, Physical Review B \textbf{92}, 094305 (2015).

\bibitem{abragam_principles_1961}
A.~Abragam, \emph{The principles of nuclear magnetism} (Oxford University
  Press, 1961).

\bibitem{bleaney_nuclear_1983}
B.~Bleaney, J.~H.~T. Pasman, and M.~R. Wells, Proceedings of the Royal Society
  of London. Series A, Mathematical and Physical Sciences \textbf{387}, 75
  (1983).

\bibitem{aminov_nuclear_1985}
L.~K. Aminov and M.~A. Teplov, Soviet Physics Uspekhi \textbf{28}, 762 (1985).

\bibitem{orbach_spin-lattice_1961}
R.~Orbach, Proceedings of the Royal Society of London. Series A. Mathematical
  and Physical Sciences \textbf{264}, 458 (1961).

\bibitem{Bottger2006}
T.~B\"ottger, C.~W. Thiel, Y.~Sun, and R.~L. Cone, Phys. Rev. B \textbf{73},
  075101 (2006).

\bibitem{Stoddart1993}
P.~R. Stoddart, P.~E. Ngoepe, P.~M. Mjwara, J.~D. Comins, and G.~A. Saunders,
  Journal of Applied Physics \textbf{73}, 7298 (1993).

\bibitem{spencer_microwave_1963}
E.~G. Spencer, R.~T. Denton, T.~B. Bateman, W.~B. Snow, and L.~G.~V. Uitert,
  Journal of Applied Physics \textbf{34}, 3059 (1963).

\bibitem{macfarlane_direct_2000}
R.~M. Macfarlane, Journal of Luminescence \textbf{85}, 181 (2000).

\end{thebibliography}
\end{document}